\documentclass[12pt]{article}
\usepackage{epsfig,amsmath,amsfonts,amssymb,amstext,afterpage,psfrag,
}
\usepackage[square,comma,sort&compress,numbers]{natbib}
\long\def\symbolfootnote[#1]#2{\begingroup%
\def\thefootnote{\fnsymbol{footnote}}\footnote[#1]{#2}\endgroup}

\def\l{\langle}
\def\r{\rangle}

\def\spose#1{\hbox to 0pt{#1\hss}}
\def\lsim{\mathrel{\spose{\lower 3pt\hbox{$\mathchar"218$}}
 \raise 2.0pt\hbox{$\mathchar"13C$}}}
\def\gsim{\mathrel{\spose{\lower 3pt\hbox{$\mathchar"218$}}
 \raise 2.0pt\hbox{$\mathchar"13E$}}}

\catcode`@=11
\def\@citex[#1]#2{%
  \if@filesw\immediate\write\@auxout{\string\citation{#2}}\fi
  \def\@citea{}\@cite{\@for\@citeb:=#2\do
    {\@citea\def\@citea{,\penalty\@m}\@ifundefined
      {b@\@citeb}{{\bf ?}\@warning
{Citation `\@citeb' on page \thepage \space undefined}}%
      \hbox{\csname b@\@citeb\endcsname}}}{#1}}
\def\citer{\@ifnextchar [{\@tempswatrue\@citexr}{\@tempswafalse\@citexr[]}}
  \def\@citexr[#1]#2{%
    \if@filesw\immediate\write\@auxout{\string\citation{#2}}\fi
    \def\@citea{}\@cite{\@for\@citeb:=#2\do
      {\@citea\def\@citea{--\penalty\@m}\@ifundefined
{b@\@citeb}{{\bf ?}\@warning
{Citation `\@citeb' on page \thepage \space undefined}}%
\hbox{\csname b@\@citeb\endcsname}}}{#1}}

\begin{document}

\begin{titlepage}

\begin{flushright}
{\small
LMU-ASC~81/13\\ 
FLAVOUR(267104)-ERC-62\\
December 2013\\
}
\end{flushright}

\vspace{0.5cm}
\begin{center}
{\Large\bf \boldmath                                               
On the Power Counting in\\     
\vspace*{0.3cm}                                                            
Effective Field Theories                     
\unboldmath}
\end{center}

\vspace{0.5cm}
\begin{center}
{\sc Gerhard Buchalla$^1$, Oscar Cat\`a$^{1,2,3}$ and Claudius Krause$^1$} 
\end{center}

\vspace*{0.4cm}

\begin{center}
$^1$Ludwig-Maximilians-Universit\"at M\"unchen, Fakult\"at f\"ur Physik,\\
Arnold Sommerfeld Center for Theoretical Physics, 
D--80333 M\"unchen, Germany\\
\vspace*{0.2cm}
$^2$TUM-IAS, Lichtenbergstr. 2a, D--85748 Garching, Germany\\
\vspace*{0.2cm}
$^3$Physik Department, TUM, D--85748 Garching, Germany
\end{center}

\vspace{1.5cm}
\begin{abstract}
\vspace{0.2cm}\noindent
We discuss the systematics of power counting in general effective
field theories, focussing on those that are nonrenormalizable
at leading order. As an illuminating example we consider 
chiral perturbation theory gauged under the electromagnetic $U(1)$ symmetry.
This theory describes the low-energy interactions of the octet
of pseudo-Goldstone bosons in QCD with photons and has been discussed
extensively in the literature. Peculiarities of the standard
approach are pointed out and it is shown how these are resolved within
our scheme. The presentation follows closely our recent discussion
of power counting for the electroweak chiral Lagrangian.
The systematics of the latter is reviewed and shown to be consistent
with the concept of chiral dimensions.
The results imply that naive dimensional analysis (NDA) is incomplete
in general effective field theories, while still reproducing the correct
counting in special cases.
\end{abstract}

\vfill

\end{titlepage}

\section{Introduction}
\label{sec:intro}

Effective field theories (EFTs) are the most efficient way of describing 
physics at a certain energy scale, provided there is a mass gap and the 
dynamical field content as well as the symmetries at that scale are known. 
What makes 
EFTs especially useful is that the operators one can build out of the fields 
can be organized according to their importance in a systematic expansion. 
The organizing principle is based on a power-counting argument. 
In weakly-coupled scenarios the power counting reduces to a dimensional 
expansion, where fields have canonical dimensions and higher-dimension 
operators are weighted with inverse powers of a cutoff scale $\Lambda$, whose 
value indicates the scale of new physics.
In this case $\Lambda$ can be arbitrarily large. 

The situation is different in spontaneously broken strongly-coupled scenarios. 
Such theories are nonrenormalizable even at leading order. As a result,
they are non-de\-coup\-ling, {\it{i.e.}}, the 
scale of new physics is no longer arbitrary but required to be at 
$\Lambda\approx 4\pi f$, where $f$ is the Goldstone-boson decay constant
of the strong sector. Correspondingly, 
strongly-coupled effective theories can only be consistent if 
based on a loop expansion, where the loop divergences at a given order are 
renormalized by operators at the following order. The EFT is predictive if 
the size of the counterterms is of the same order as the loop contributions, 
to which they are related by renormalization~\cite{Manohar:1983md}. 
Power counting is no longer based on 
the canonical dimension of fields and should be constructed instead by 
analysing the loop structure of a given (leading-order) Lagrangian. 
Knowledge of the effective Lagrangian at leading order is therefore necessary. 

This strategy for the construction of EFTs with strongly-coupled dynamics 
is notably simplified in specific cases. The paradigm of simplicity is 
chiral perturbation theory ($\chi$PT)~\cite{Weinberg:1978kz,Leutwyler:1993iq} 
for massless pions, where the power counting reduces to an expansion in 
derivatives. 
When external sources are added and pion masses switched 
on~\cite{Gasser:1983yg,Gasser:1984gg}, chiral symmetry is explicitly broken. 
It is common to extend the derivative counting to these new objects too. 
This formal assignment of derivative counting to couplings and fields 
goes under the name of chiral dimensional counting ($\chi$DC). 
It is constrained by the requirement that the terms in the leading-order 
Lagrangian must have the same chiral dimension. Following this method, 
extensions of $\chi$PT to include dynamical photons~\cite{Urech:1994hd} and 
leptons~\cite{Knecht:1999ag} have been formulated. 

Defined in this way, the assignment of chiral dimensions seems
unsatisfactory in some respects.
First, chiral dimensions may suggest a misleading interpretation of 
the physics of strongly-coupled dynamics. For instance, the electromagnetic
coupling $e$ is a parameter independent of chiral symmetry breaking, yet 
it has an assigned momentum scaling. $\chi$DC should thus be rather 
understood as a formal tool. However,
to the best of our knowledge, $\chi$DC has never been justified in terms of 
a diagrammatic power counting. Second, $\chi$DC alone does not yet
allow one to construct an operator basis. 
In other words, it is no substitute for a full-fledged power-counting formula. 
These points have led to some confusion in the literature, especially in 
studies of electroweak effective theories~\cite{Nyffeler:1999ap,Hirn:2004ze}.  
  
An alternative approach is naive dimensional analysis 
(NDA)~\cite{Manohar:1983md}. Using the chiral quark model as a paradigmatic
example, a simple set of rules has been inferred to build power-counting 
formulas for generic EFTs. In a nutshell, fundamental and composite fields are 
simply associated with different scales, $1/\Lambda$ for the former and 
$1/f$ for 
the latter~\cite{Georgi:1992dw}. This prescription is in contrast to $\chi$DC. 

Both $\chi$DC and NDA have the common objective to describe the systematics 
of EFTs in which strongly-coupled and weakly-coupled sectors mix. Since both 
methods rely on some sort of dimensional expansion encoded in a set of rules, 
it would be interesting to explore the relation between both approaches and 
understand whether they are mutually consistent. In this paper we will 
clarify these issues by reassessing the $\chi$DC and NDA prescriptions in the 
light of a general power-counting formula for strongly-coupled theories with 
fermions, gauge bosons and scalars, initially derived 
in~\cite{Buchalla:2012qq,Buchalla:2013rka}. We will specialize it to the 
strong and electroweak interactions and compare it with the predictions of 
$\chi$DC and NDA. 

We show that $\chi$DC is a consistent prescription and can be rephrased in 
terms of systematic power-counting arguments. Its formal and, strictly
speaking, unphysical scaling rules turn out to be the price to pay in order 
to force a simple dimensional counting onto a strongly-coupled EFT. 
We also show that the rules of NDA are not valid in general and lead to 
contradictions, for instance in the electroweak interactions. We point out 
how they should be modified. In particular, we will find out that power 
counting is insensitive to the fundamental or composite nature of fermions, 
yet very sensitive to their couplings with the Goldstone modes. 

This paper is organized as follows. We revisit $\chi$PT with 
dynamical photons in Section~\ref{sec:chipt} and derive the relevant 
power-counting formula. The latter is a new result, in spite of the fact 
that this EFT has been widely used. The formula allows us to prove that 
the definition of chiral dimensions employed in~\cite{Urech:1994hd} is 
both consistent and unique. In Section~\ref{sec:ewchil} we 
turn to the electroweak interactions and discuss the  
power counting that applies when
the spontaneous symmetry breaking is induced by 
strongly-coupled dynamics. We show how the general results derived 
in~\cite{Buchalla:2012qq,Buchalla:2013rka} can be reinterpreted in the 
language of chiral dimensions.
Section~\ref{sec:nda} comments on the implications for
the NDA prescription.    
We conclude in Section~\ref{sec:concl}.

\section{Chiral perturbation theory with photons}
\label{sec:chipt}

\subsection{Lagrangian at leading order}
\label{subsec:llo}

Many of the essential features in the power counting of 
strongly-coupled effective field theories are already present
in the case of chiral perturbation theory of pions and kaons
coupled to electromagnetism. Due to its relative simplicity
this case will serve as an instructive example for our discussion.
 
Under $SU(3)_L\times SU(3)_R$ the Goldstone boson matrix $U$ transforms as
\begin{equation}\label{uhglgr}
U\rightarrow g_L U g^\dagger_R,\qquad g_{L,R}\in SU(3)_{L,R}
\end{equation}
The explicit relation between the 
matrix $U$ and the Goldstone fields $\varphi^a$ is
\begin{equation}\label{uudef}
U=\exp(2i\Phi/f),\qquad
\Phi=\varphi^a T^a
\end{equation}
where $T^a=T_a=\lambda^a/2$ are the generators of $SU(3)$
and $f\approx 93\, {\rm MeV}$ is the Goldstone-boson decay constant.

The vectorial subgroup of $SU(3)_L\times SU(3)_R$ is gauged under the
electromagnetic $U(1)$, so that the covariant derivative is given by
\begin{equation}\label{dcovu}
D_\mu U=\partial_\mu U+i e A_\mu [Q,U]
\end{equation}
where $Q={\rm diag}(2/3,-1/3,-1/3)$. 

The full chiral symmetry $SU(3)_L\times SU(3)_R$ is broken
by the quark-mass term ($\chi$) and by electromagnetism ($Q$). 
This can be implemented in the standard way through the
corresponding spurions transforming as
\begin{equation}\label{chiq}
\chi \to g_L \chi g^\dagger_R ,\qquad  Q_L\to g_L Q_L g^\dagger_L,
\qquad  Q_R\to g_R Q_R g^\dagger_R 
\end{equation} 
with the identification $Q_L=Q_R=Q$.
Similarly, $\chi=2 B {\cal M}$ with ${\cal M}={\rm diag}(m_u,m_d,m_s)$.

The leading-order Lagrangian can then be written as 
\cite{Urech:1994hd,Knecht:1999ag,Neufeld:1995mu,Meissner:1997fa,Knecht:1997jw,Ecker:2000zr}
\begin{equation}\label{llo}
{\cal L}_{LO} = \frac{f^2}{4}\ \l D_\mu U^\dagger D^\mu U\r\, 
+\frac{f^2}{4}\ \l U^\dagger \chi + \chi^\dagger U\r\,
-\frac{1}{4} F_{\mu\nu} F^{\mu\nu} + e^2 \Delta \l U^\dagger Q U Q\r
\end{equation}
where $\l\ldots\r$ denotes the trace.

Eq. (\ref{llo}) is the lowest-order approximation to the theory
of pseudo-Goldstone bosons and photons with typical energies of
the order of $f$. The expansion parameter governing higher-order
corrections is
$f^2/\Lambda^2$, with $\Lambda=4\pi f$ the scale of chiral symmetry
breaking. All terms in (\ref{llo}) are indeed of leading order ($\sim f^4$)
in this expansion. This follows from the fact that in a 
scattering process involving Goldstone bosons and photons at a typical energy
$\sim f\ll\Lambda$, the relevant quantities scale as
\begin{equation}\label{scaling}
\partial_\mu\sim f,\quad \varphi^a\sim f,
\quad \chi\sim f^2,\quad A_\mu\sim f,\quad e\sim 1 
\end{equation}
The spurion $\chi$ is proportional to the pseudo-Goldstone masses squared,
which are counted in the standard way as $\sim f^2$, consistent with
the homogeneous scaling of the meson propagator.
The coupling $e$ is an independent parameter, which can be viewed as a 
quantity of order one. A further expansion for $e^2\ll 1$
can always be performed if desired. 

The scaling $\sim f^4$ follows immediately for the first three terms 
in (\ref{llo}).  
The last term has no derivatives and amounts to a potential for the
Goldstone bosons, induced by virtual photons. 
It is proportional to the cut-off squared, but
carries a loop suppression. It scales as \cite{Das:1967it} 
\begin{equation}\label{deltaf4}
\Delta\sim f^2 \frac{\Lambda^2}{16\pi^2}\sim f^4
\end{equation}
and is consistently included in ${\cal L}_{LO}$.
We note that the leading-order Lagrangian has terms
with canonical dimension zero (second and fourth term), two (first term) and 
four (third term). As is well known,
the Lagrangian is not ordered by canonical dimension of operators
in the case of a strongly-interacting system. This is hardly suprising,
since already the first term in ${\cal L}_{LO}$ contains operators
of arbitrarily large canonical dimension when expanded out in powers of
the field $\Phi$. 

Rather than by dimensional counting, the higher-order terms correcting 
the Lagrangian in (\ref{llo}) are governed by a loop expansion, which
corresponds to a series in powers of $1/(16\pi^2)=f^2/\Lambda^2$. 
The systematics of this construction can be described by
a power-counting formula, which we discuss in the following section.

\subsection{Power counting and the Lagrangian at NLO}
\label{subsec:powc}

The leading-order Lagrangian (\ref{llo}) is nonrenormalizable. 
Corrections can be organized in the form of a loop expansion. 
The relevant power counting has been discussed in
\cite{Buchalla:2012qq,Buchalla:2013rka} for the electroweak chiral Lagrangian.
It makes use of the assumption that the loop effects in the strong sector
$\sim 1/(16\pi^2)$ are actually of the same order of magnitude as the
corresponding coefficients of NLO operators $\sim f^2/\Lambda^2$
\cite{Manohar:1983md,Georgi:1992dw}. This implies the identification
$\Lambda=4\pi f$. A priori, the coefficients $\sim f^2/\Lambda^2$ need only be
at least of the size of the loop contribution $\sim 1/(16\pi^2)$, giving 
$\Lambda\lsim 4\pi f$. The approximate equality is a natural assumption for
QCD. In the case of electroweak symmetry breaking
the assumption is also justified as long as new-physics states appear
only at the (few) TeV scale or above. 

Adapted to the present case, the power counting for a 
diagram ${\cal D}$ with $L$ loops, built from the vertices of (\ref{llo}),
can be summarized by the formula
\begin{equation}\label{pcchpta}               
{\cal D}\sim  
\frac{p^d}{\Lambda^{2L}}\  
\left(\frac{F_{\mu\nu}}{f}\right)^V\  \left(\frac{\varphi}{f}\right)^B   
\end{equation}
where the power of external momenta $p$ is
\begin{equation}\label{powerd1}                                            
d\equiv 2L+2-V-m-2r-2\omega_\chi - 2\omega_Q                   
\end{equation}
Here $V$ is the number of external factors of photon 
field strength $F_{\mu\nu}$, $m$ ($r$) is the number of photon-meson
vertices of the form $A_\mu\phi^l$ ($A^2_\mu\phi^s$), and
$\omega_\chi$ ($\omega_Q$) is the number of meson vertices
from the term with $\chi$ ($Q$) in (\ref{llo}). 
The number $B$ of external meson lines does not enter $d$ in (\ref{powerd1}). 

Assuming dimensional regularization,
an exponent $d\geq 0$ in (\ref{pcchpta}) indicates a divergence by
power counting, as well as the number of derivatives in the corresponding
counterterm (not counting those in the factors of $F_{\mu\nu}$). 
Using (\ref{powerd1}), one finds that the classes of
counterterms at next-to-leading order ($L=1$) are exhausted by the
six cases
\begin{equation}\label{classnlo}
(V,\omega_\chi+\omega_Q;d):\quad
(0,0;4),\, (0,1;2),\, (0,2;0),\, (1,0;2),\, (1,1;0),\, (2,0;0)
\end{equation} 
The explicit operators in each of these classes that are compatible
with chiral symmetry and its breaking by spurions, and even under
$C$ and $P$, can be listed as follows.

{\bf (0,0;4):}
\begin{align}\label{c004}
& \l D_\mu U^\dagger D^\mu U\r^2,\quad 
\l D_\mu U^\dagger D_\nu U\r\, \l D^\mu U^\dagger D^\nu U\r,\quad
\l D_\mu U^\dagger D^\mu U D_\nu U^\dagger D^\nu U\r
\end{align}

{\bf (0,1;2):}
\begin{align}\label{c012}
& \l D_\mu U^\dagger D^\mu U\r\, \l \chi U^\dagger + U\chi^\dagger \r ,\quad 
\l D_\mu U^\dagger D^\mu U (U^\dagger\chi + \chi^\dagger U)\r\nonumber\\
& e^2 \l D_\mu U^\dagger D^\mu U\r \l U^\dagger Q U Q\r ,\quad
e^2 \l D_\mu U^\dagger(UQ\pm QU)\r^2\nonumber\\
& e^2 \l (D_\mu U^\dagger D^\mu U + D_\mu U D^\mu U^\dagger)Q\r ,\quad
e^2 \l D_\mu U^\dagger D^\mu U (U^\dagger QUQ+Q U^\dagger QU)\r
\end{align}

{\bf (0,2;0):}
\begin{align}\label{c020}
& \l \chi U^\dagger \pm U\chi^\dagger\r^2,\quad
\l \chi U^\dagger \chi U^\dagger + U\chi^\dagger U\chi^\dagger\r \nonumber\\
& e^2 \l \chi U^\dagger + U\chi^\dagger\r \l U^\dagger Q U Q\r,\quad
e^2 \l (\chi U^\dagger + U\chi^\dagger+U^\dagger \chi+\chi^\dagger U)Q\r\nonumber\\
&e^2 \l (\chi U^\dagger \pm U\chi^\dagger)QUQ U^\dagger +
    (\chi^\dagger U \pm U^\dagger \chi)Q U^\dagger QU\r ,\quad
e^4 \l U^\dagger Q U Q\r^2 
\end{align}

{\bf (2,0;0):}
\begin{align}\label{c200}
& e^2 F_{\mu\nu} F^{\mu\nu}\, \l U^\dagger Q U Q\r
\end{align}

Classes {\bf (1,0;2)} and {\bf (1,1;0)} contain no independent operators.

The complete list of NLO operators in (\ref{c004}) -- (\ref{c200})
is consistent with the results originally obtained in \cite{Urech:1994hd}.
NLO counterterms that reduce to LO structures when 
$Q_{L,R}\to Q={\rm const.}$ are not included in our list above.

\subsection{Alternative scheme based on chiral dimensions}
\label{subsec:chidim}

It is possible to interpret the results of the previous sections
in a somewhat different, but equivalent, way.
Since the chiral Lagrangian is organized as a loop expansion,
the different orders are just given by $L$, the number of loops.
For convenience we may define $2L+2$
as the {\it chiral order\/} of the corresponding terms in the Lagrangian.  
Using (\ref{powerd1}) the chiral order can be written as
\begin{equation}\label{chpta2lp2}
2L+2 = d + V + m + 2r + 2\omega_Q + 2\omega_\chi
\end{equation}
This implies that the chiral order $2L+2$ is obtained for any term
by adding the number $d$ of derivatives it contains (not counting 
those in $F_{\mu\nu}$), the number $V$ of photon field-strength factors
$F_{\mu\nu}$, the total number $m+2r+2\omega_Q$ of couplings $e$, and
twice the number $\omega_\chi$ of factors of $\chi$.   
A {\it chiral dimension\/} $[x]_c$ can therefore be assigned to any
quantity $x$ in the chiral Lagrangian according to the number it
contributes to the chiral order of an operator. It follows from
(\ref{chpta2lp2}) that
\begin{equation}\label{chidim1}
[\partial_\mu]_c =1,\quad [F_{\mu\nu}]_c=1,\quad [e]_c=1,
\quad [\chi]_c = 2,\quad [U]_c=0
\end{equation}
and consequently 
\begin{equation}\label{chidim2}                                            
[A_\mu]_c =0,\quad [D_\mu]_c=1,\quad [\Phi]_c=0          
\end{equation}
Applying these rules, we see that all terms in the leading-order Lagrangian
(\ref{llo}) have chiral order 2, irrespective of their canonical
dimension. All the terms at NLO at the end of Section \ref{subsec:powc}
have chiral order 4. We also note the well-known result that in the case of 
pure chiral perturbation theory  (without photons and $\chi$ term) the chiral
order is simply given by the number of derivatives, $2L+2=d$, which
follows from (\ref{chpta2lp2}) as a special case.
Based on this result chiral perturbation theory is often viewed 
as an expansion in the number of derivatives. While this is correct
in the simplest case, it should not be misinterpreted as being
equivalent to an expansion in (canonical) dimension.   

The assignment in (\ref{chidim1}) and (\ref{chidim2}) is the counting 
introduced for chiral perturbation theory with photons by Urech 
in \cite{Urech:1994hd}. The justification given there was somewhat
different from ours, although essentially equivalent in the end.
It was noted in \cite{Urech:1994hd} that the formal counting in
(\ref{chidim1}) and (\ref{chidim2}) leads to a homogeneous chiral order
of 2 for the Lagrangian (\ref{llo}). The one-loop counterterms were then
computed by the heat-kernel method and shown to correspond to the
terms of chiral order 4. We emphasize, however, that the assignment 
for $e$ and $A_\mu$ in 
(\ref{chidim1}) and (\ref{chidim2}) is purely formal and has a consistent
basis in the power-counting formulas of (\ref{powerd1}) and
(\ref{chpta2lp2}). Those can be derived from the physical scaling
of parameters and fields and a priori without recourse to the
concept of chiral dimension.


\section{Electroweak chiral Lagrangian}
\label{sec:ewchil}

\subsection{Basic structure}
\label{subsec:basic}

The leading-order electroweak chiral Lagrangian including 
a light Higgs boson can be schematically written in the 
form \cite{Grinstein:2007iv,Contino:2010mh,Azatov:2012bz}
\begin{align}\label{ewlo}
{\cal L}_{EW,LO}= &-\frac{1}{4}X_{\mu\nu} X^{\mu\nu} +\bar\psi i\not\!\! D\psi
+\frac{v^2}{4}\l D_\mu U^\dagger D^\mu U\r\, (1+F(h)) 
-y v[\bar\psi_L F_Y(h)U\psi_R +{\rm h.c.}] \nonumber\\
&+\frac{1}{2}\partial_\mu h \partial^\mu h - V(h) 
\end{align}
The construction of next-to-leading order terms (in the higgsless case)
dates back to the work of 
\cite{Appelquist:1980vg,Longhitano:1980iz,Appelquist:1984rr}.
Including a light Higgs, a subset of NLO operators has been given
in \cite{Alonso:2012px}, the power counting and the complete list has been 
worked out in \cite{Buchalla:2012qq,Buchalla:2013rka}.
Beyond power counting, the actual size of the operator coefficients may
be further suppressed through additional symmetries (such as flavour or CP).

In \cite{Buchalla:2012qq,Buchalla:2013rka} it has been shown that the result 
of power counting for a generic $L$-loop diagram ${\cal D}$ built from 
vertices of the leading-order Lagrangian (\ref{ewlo}) can be expressed as 
\begin{equation}\label{pcupsix}
{\cal D}\sim 
v^{2+2\omega}(yv)^\nu (gv)^{\gamma}
\frac{p^d}{\Lambda^{2L}}\
\left(\frac{\psi}{v}\right)^F
\left(\frac{X_{\mu\nu}}{v}\right)^V\  \left(\frac{\varphi}{v}\right)^B\
\left(\frac{h}{v}\right)^H
\end{equation}
where the power of external momenta $p$ is
\begin{equation}\label{powerd}
d\equiv 2L+2-\frac{F}{2}-V-\nu-\gamma-2\omega
\end{equation}
Here $F$ and $V$ are the number of external fermions and 
gauge-boson field-strength factors, respectively.
$\nu$ is the number of Yukawa couplings, $\gamma$ the number of
gauge couplings, and $\omega$ the number of (non-derivative) Higgs-boson 
self interactions.

\subsection{Chiral dimensions}
\label{subsec:ewchd}

We note that this general result, which includes chiral fermions,
can also be interpreted using the concept of chiral dimensions.
This leads to an immediate generalization of the case discussed in 
Section \ref{subsec:chidim}. Indeed, the chiral order may now
be written as
\begin{equation}\label{ewchidim}
2L+2= d+\frac{F}{2}+V+\nu+\gamma+2\omega
\end{equation}
which corresponds to the following assignment of chiral dimension
to derivatives $\partial_\mu$, Goldstone fields $\varphi$, Higgs fields $h$,
gauge-field strengths $X_{\mu\nu}$, fermions $\psi_{L,R}$, gauge couplings $g$
and Yukawa couplings $y$:
\begin{equation}\label{chidim3}
[\partial_\mu]_c=1,\quad [\varphi]_c=[h]_c=0,\quad
[X_{\mu\nu}]_c=1,\quad [\psi_{L,R}]_c=\frac{1}{2},\quad [g]_c=[y]_c=1
\end{equation}
With this counting all the terms in the leading-order Lagrangian
(\ref{ewlo}) have chiral dimension 2. The potential is assumed to
be radiatively generated and thus implicitly includes a factor
of $g^2$ or $y^2$.

The classes of next-to-leading order operators worked out in
\cite{Buchalla:2012qq,Buchalla:2013rka}
\begin{equation}\label{nloclass}
UhD^4,\quad g^2X^2Uh,\quad gXUhD^2,\quad 
y^2\psi^2UhD,\quad y\psi^2 UhD^2,\quad y^2\psi^4 Uh
\end{equation}
have chiral dimension 4. 
Note that the counting based on chiral dimension necessitates the
inclusion of the appropriate powers of couplings in the NLO
counterterms.

The chiral dimensions in (\ref{chidim3}) have been discussed previously
in the work of Nyffeler and Schenk \cite{Nyffeler:1999ap} in the context
of the (higgsless) electroweak chiral Lagrangian.
However, the assignment of chiral dimensions in \cite{Nyffeler:1999ap}
was not based on an explicit power-counting analysis of loop
corrections and counterterms. Instead, \cite{Nyffeler:1999ap}
defined the chiral dimensions of couplings and fields such as to ensure 
a homogeneous scaling of all the kinetic terms, following \cite{Urech:1994hd}.
For instance, with $[D_\mu U^\dagger D^\mu U]_c =2$, the same scaling of the
fermion kinetic term, $[\bar\psi i\!\not\!\! D\psi]_c=2$, is obtained for
$[\psi]_c=1/2$, while $[y\bar\psi_L U\psi_R]_c=2$ then requires $[y]_c=1$, 
{\it etc.} This leads to a systematic assignment of chiral dimension $2L+2$ 
to terms of order $L$ in the effective-theory series, in agreement with
our derivation above.

Nevertheless, the mere assignment of chiral dimensions as in (\ref{chidim3})
does not by itself suffice to specify the full systematics of the
effective-theory power counting. A clear example is given by
the 4-fermion operators $\bar\psi\psi\,\bar\psi\psi$.
These terms have a chiral dimension of 2 and would seem to be part
of the leading-order Lagrangian. However, such an assignment would not be 
consistent for the electroweak chiral Lagrangian.
A 4-fermion operator would arise at leading-order
for instance from the exchange of a heavy resonance of mass $\Lambda=4\pi v$
with a strong coupling $\sim 4\pi$ to the fermionic current $\bar\psi\psi$,
giving an unsuppressed coefficient $\sim (4\pi)^2/\Lambda^2=1/v^2$ to the 
4-fermion term. Even for the top quark, which has the strongest coupling to the
symmetry-breaking sector of any standard-model fermion, this coupling is 
only of order 1, giving a suppression of the coefficient. 
Assuming therefore a weak coupling of fermions to the strong sector
(of order unity or less), eliminates 4-fermion terms from the
leading-order Lagrangian.
We also note that in the case of a strong coupling of a fermion to
the symmetry-breaking sector, its mass would be $\sim 4\pi v=\Lambda$.
Such a fermion would not be part of the spectrum at low energies and
therefore not included as a field in the effective Lagrangian.

Similar arguments suggest that any operator of the form $X^2Uh$,
which also has a chiral dimension of 2, cannot appear at leading
order since the gauge fields $X$ are again weakly coupled to the strong
sector, that is with couplings of order 1 instead of $4\pi$.
Note that an operator $X^2Uh$ appearing at leading order could induce,
for instance, $h\to\gamma\gamma$ decays with an amplitude larger
than in the standard model by a factor of $16\pi^2$, which is
excluded by experiment.

Rather than at leading order, terms such as $\bar\psi\psi\,\bar\psi\psi$ or 
$X^2Uh$ arise at next-to-leading order where they come
with explicit factors of couplings $y^2$ or $g^2$, which shifts their
chiral dimension to 4.

In \cite{Nyffeler:1999ap} 4-fermion operators have been listed as
leading-order terms, on the formal ground that they have chiral
dimension 2. Even though they seem to have been recognized as
phenomenologically undesirable at this order, the physical implications 
were not clearly spelled out. 
Similar issues apparently prompted \cite{Hirn:2004ze} to question
the standard assignment of chiral dimensions and to attempt a
modification that employed spurions to eliminate unwanted terms.  
This approach remained largely inconclusive.

The analysis discussed in the present article addresses both the physical 
content of the electroweak chiral Lagrangian and the justification of chiral 
dimensions in terms of standard methods of power counting. This clarifies the 
systematics of the effective Lagrangian and resolves the peculiarities 
encountered in \cite{Nyffeler:1999ap,Hirn:2004ze}, and more recently
in the discussion of derivative counting in \cite{Alonso:2012px}.

\subsection{Counting of chiral dimensions to all orders}
\label{subsec:chdall}

In the previous sections we have discussed the power counting to all
orders in the loop expansion, but only with vertices of the
leading-order Lagrangian. This is sufficient to construct the
counterterms at all orders.
Here we show how the counting of chiral dimensions is extended to the
fully general case, including any loop order, as well as vertices from
any order in the effective Lagrangian.

A general term in the effective Lagrangian can be denoted by
\begin{equation}\label{termi}
\kappa^{k_i} \psi^{F_i} X_{\mu\nu}^{V_i} D^{d_i}\, Uh
\end{equation}
with a fixed number $k_i$ of couplings $\kappa$ (gauge or Yukawa couplings),
$F_i$ fermion fields $\psi$, $V_i$ field-strength factors $X_{\mu\nu}$,
$d_i$ covariant derivatives $D$, and an arbitrary number of
(pseudo-) Goldstone bosons ($Uh$).
This term defines vertices of type $ij$, in general with $B_{ij}$
Goldstone lines. The total chiral dimension of (\ref{termi}) determines the
loop order $L_i$, that is the order of the term in the effective theory, 
from (\ref{ewchidim}) as
\begin{equation}\label{chidimli}
d_i+k_i+\frac{F_i}{2}+V_i = 2 L_i + 2 
\end{equation}

We consider next an arbitrary diagram ${\cal D}$ with $L$ loops and
any number of type-$ij$ vertex insertions. Denoting the number of external 
(internal) fermion, gauge-field and Goldstone lines
by $F$, $V$, $B$ (${\cal F}$, ${\cal V}$, ${\cal B}$), respectively,
the familiar topological identities give
\begin{align}\label{topid}
F+ 2{\cal F} &= \sum_i n_i F_i\nonumber\\
V+ 2{\cal V} &= \sum_i n_i V_i\nonumber\\
B+ 2{\cal B} &= \sum_{i,j} n_{ij} B_{ij}\nonumber\\
L &= {\cal F} + {\cal V} + {\cal B} -\sum_{i,j} n_{ij} + 1
\end{align} 
where $n_{ij}$ is the number of vertices of type $ij$ and
$n_i=\sum_j n_{ij}$.

Defining the total number of couplings in diagram ${\cal D}$ by
$k=\sum_i n_i k_i$, one finds from (\ref{topid})
\begin{equation}\label{chidimgen}
d+k+\frac{F}{2}+V = 2L + 2 + \sum_i n_i\, 2 L_i
\end{equation}
This relates the chiral dimension of diagram ${\cal D}$ on the left to 
the number of loops $L$ and the loop order of the vertex insertions $L_i$.
Put differently, the chiral dimension $[{\cal D}]_c=d+k+F/2+V$ gives the
total loop order as
\begin{equation}\label{chidimloop}
\frac{[{\cal D}]_c -2}{2} = L + \sum_i n_i L_i
\end{equation} 
This implies that, in general, the chiral dimension of a diagram simply
counts the number of loops. More precisely, it gives the order
of a diagram, or an operator, in the loop expansion, on which the
effective theory construction is based. 

It is interesting to compare the systematics of chiral dimensions
with the standard counting by canonical dimension that governs the
low-energy description of weakly-coupled theories.
In the latter case, the leading-order, dimension-4 Lagrangian is 
renormalizable and higher-dimensional terms can be added as corrections.
They are increasingly suppressed by inverse powers $1/M^{d-4}$ of the
new-physics scale $M$ with increasing canonical dimension $d$
of the operators. As is well known, in a general diagram,
with an arbitrary number of operator insertions, the corresponding powers
of $1/M$ simply add up to the total power of $1/M$ for the entire
diagram, independently of the number of loops.
Eq. (\ref{chidimloop}) implies that a formally similar rule holds
for the counting of chiral dimensions. However, as discussed in
the previous section, collecting the terms of a given chiral dimension
is not sufficient to establish the operators at a given order in
the strongly-coupled case. A consistent definition of the
leading-order Lagrangian and an analysis of counterterms is also needed.
This important difference to the dimensional case is a consequence
of the fact that the theory is organized as a loop expansion.

\section{Naive dimensional analysis}
\label{sec:nda}

The order at which a given operator appears in a general effective
field theory is often determined using naive dimensional analysis (NDA).
This procedure has been introduced in \cite{Manohar:1983md} in the
context of the chiral quark model, with the understanding that its
validity is more general. In the example of the chiral quark model
with quarks $\psi$, gluons $G_\mu$, gauge coupling $g$, Goldstone bosons
$\varphi$ and derivatives $p$, the coefficient of a general term in the
effective Lagrangian has been given in \cite{Manohar:1983md} as
\begin{equation}\label{ndamg}
\left(\frac{\varphi}{f}\right)^A\, 
\left(\frac{\psi}{f\sqrt{\Lambda}}\right)^B\,
\left(\frac{g G_\mu}{\Lambda}\right)^C\,
\left(\frac{p}{\Lambda}\right)^D\, f^2\Lambda^2
\end{equation}
Eq.~(\ref{ndamg}) can be easily translated to the language of 
chiral dimensions. The order of suppression of the different terms in the 
effective theory is determined by the inverse powers of $\Lambda$ it contains.
Since $(f^2/\Lambda^2)^L$ corresponds to the loop order,
(\ref{ndamg}) implies
\begin{equation}\label{nda2}
2L+2=D+\frac{B}{2}+C
\end{equation}
or, in the notation of (\ref{ewchidim}),
\begin{equation}\label{nda3}
2L+2=d+\frac{F}{2}+V+\gamma
\end{equation}
with the definitions $d\equiv D-C$, $F\equiv B$, $V\equiv C$,
and with the additional assumptions $\gamma=C=V$, $\nu=\omega=0$. 
Differences arise since in (\ref{ndamg}) the number of gauge fields ($V$) and 
of gauge couplings ($\gamma$) are identified (because internal gauge-boson 
lines are neglected), and Yukawa terms ($\nu$)
and Goldstone-boson non-derivative couplings are not included.

The formula (\ref{nda3}) agrees with the result (\ref{ewchidim}) in
assigning the correct chiral dimension to $\varphi$
(chiral dimension $0$), $\psi$ ($1/2$), $gG_\mu$ ($1$) and $p$ ($1$).
Note in particular that the chiral dimension of the vector-like fermion
$\psi$ here is identical to the chiral dimension of the chiral fermions
in (\ref{chidim3}). However, the NDA prescription does not
specify the separate counting of $g$ and $G_\mu$. As we have seen,
the assignment of chiral dimension has to be $[g]_c=1$ and $[G_\mu]_c=0$
to achieve a universal counting for parameters and fields.
Such a universal counting is an essential objective of NDA.

The counting of NDA has been summarized in a compact way in 
\cite{Georgi:1992dw}. This paper states that the size of the coefficient
for any term in the effective Lagrangian is obtained by 
including an overall factor of $f^2\Lambda^2$, a factor of $1/f$
for each strongly-interacting field, and factors of $\Lambda$ to get the
dimension to 4. Weakly interacting fields enter with a suppression by
inverse powers of $\Lambda$ according to their canonical dimension.
Implicit is the assumption that gauge couplings $g$ are to be treated
as factors of order one.
These rules have been abstracted from (\ref{ndamg}) and postulated to
be of general validity for effective theories with Goldstone bosons from
a strongly-interacting sector.

We point out that the rules of NDA used in this form are incomplete
and can lead to incorrect results. 
As an illustration let us consider the following examples:
\begin{itemize}
\item 
The coefficient of the photon kinetic term in (\ref{llo})
is obtained in NDA as $f^2\Lambda^2/\Lambda^4=f^2/\Lambda^2$
instead of the correct size of order 1. 
This problem had already been noted in \cite{Manohar:1983md},
but the consequences for NDA were not fully explored.
\item
Applying the NDA rules to the electromagnetic mass term
in (\ref{llo}), one finds a coefficient of $f^2\Lambda^2$
instead of $f^4$. In this case the coefficient is too large
by a factor of $\Lambda^2/f^2$.
\item 
In the electroweak chiral Lagrangian the operator classes $\psi^2 UhD$ 
and $\psi^2 UhD^2$ are both present at next-to-leading order. 
It is clear that no scaling of fields and derivatives can be devised 
that would yield coefficients of the same order for these two classes.
The NDA scaling in (\ref{ndamg}) would suggest coefficients
of order 1 for the first class and of order $1/\Lambda$ for the second,
different from the correct scaling $\sim 1/\Lambda^2$ in
both cases. As discussed above, an appropriate 
(formal) scaling of the associated Yukawa couplings will
ensure the correct size of the coefficients. 
\end{itemize}

In summary, while we agree with the basic assumptions of NDA,
which have been clearly explained in \cite{Manohar:1983md,Georgi:1992dw},
we find that the specific counting rules given in these papers are
incomplete. Complete and consistent formulations of the power counting
are described in Sections \ref{sec:chipt} and \ref{sec:ewchil}. 
In particular, a direct comparison of (\ref{ndamg}) with the power counting 
in (\ref{pcupsix}) shows that agreement is reached if the NDA formula is 
enlarged by a factor
\begin{align}\label{pref}
R=\left(\frac{1}{4\pi}\right)^{2\omega}\left(\frac{y}{4\pi}\right)^{\nu}
\left(\frac{g}{4\pi}\right)^{\gamma-V}
\end{align}

A similar generalization of the NDA formula has recently been
considered in \cite{Jenkins:2013sda}. There it has been used
in the context of weakly-coupled effective field theories with dimensional
counting, rather than for strongly-coupled scenarios.
The main result of \cite{Jenkins:2013sda} is an
identity relating the perturbative order of a diagram (i.e. the
total number $2N$ of weak couplings) to the number of loops $L$ as
\begin{equation}\label{nlw} 
N=L+w-\sum_k w_k
\end{equation}
where $w$ and $w_k$ are the NDA weights, defined in \cite{Jenkins:2013sda},
of the diagram and the inserted operators, respectively. We find that 
in general the weight $w$ can be written in terms of the canonical dimension 
$[{\cal O}]$ and the chiral dimension $[{\cal O}]_c$ of an operator as 
\begin{equation}\label{ndaw}
w\equiv\frac{[{\cal O}]-[{\cal O}]_c}{2}-1
\end{equation}
The identity (\ref{nlw}) can be immediately obtained from
(\ref{topid}) and is therefore seen to be entirely of topological
origin. As such it does not by itself determine the degree of divergence
or the order of a given operator in the EFT expansion.
This is reflected in the fact that the identity holds both
for the chiral Lagrangian and the Lagrangian with dimensional counting,
which have a very different organization of the EFT counting.

\section{Conclusions}
\label{sec:concl}

We have presented a detailed discussion of the power counting for
effective field theories valid at a scale $v$ and with strong dynamics
at their cut-off $\Lambda\gg v$.
This counting is the key element for organizing the possible terms in the
effective Lagrangian according to their order in powers of
$v^2/\Lambda^2$. The basic assumptions can be stated as follows:
\begin{itemize}
\item
The degrees of freedom at the low scale $v$ are, in general,
(chiral) fermions $\psi_{L,R}$, gauge fields $A_\mu$ and (pseudo-)
Goldstone bosons $\varphi$.
\item
A mass gap separates the scale $v$ from the high scale $\Lambda$,
which has been integrated out in the low-energy effective theory.
\item
At the scale $\Lambda$ (part of) the dynamics is strongly coupled,
the natural cut-off is then $\Lambda=4\pi v\gg v$.
\item
The Goldstone sector is strongly coupled, with couplings $\sim 4\pi$,
to the strong interactions at $\Lambda$.
\item
Chiral fermions and gauge fields are weakly coupled to the dynamics at
$\Lambda$, that is with couplings of order unity (or smaller).
\end{itemize}
Important examples for such a scenario are the chiral perturbation theory
for pions and kaons in the presence of electromagnetism, or the 
electroweak chiral Lagrangian with a light (pseudo-Goldstone) Higgs. 
In the latter case the actual cut-off may be at a scale $4\pi f$, with $f > v$.
When the parameter $\xi\equiv v^2/f^2$ is taken to zero, the ordinary 
standard model is recovered. Expanded to first order in $\xi$, 
the electroweak chiral Lagrangian contains \cite{Buchalla:2013rka} 
the SILH framework \cite{Giudice:2007fh}.
The full chiral Lagrangian amounts to a resummation of terms to all
orders in $\xi$, which is parametrically viewed as a quantity of order one.   

We have emphasized the importance of specifying the leading-order
Lagrangian consistently with the assumptions above.
The leading-order Lagrangian provides the basis for the
power-counting analysis of loop corrections and their divergence
structure. The latter determines the required classes of counterterms, which
yield the higher-order operators in the effective Lagrangian.

Previous treatments of power counting appear to have followed one of
two different lines of approach, the first employing naive dimensional 
analysis \cite{Manohar:1983md,Georgi:1992dw}, the second using the concept 
of chiral dimensions \cite{Urech:1994hd,Nyffeler:1999ap,Hirn:2004ze}.
There seems to have been little overlap between the parts of the literature 
applying one or the other framework.
We have shown how the two methods are related. In particular,
we have demonstrated how chiral dimensions follow from standard
power counting and we have clarified the physical assumptions
that are needed in addition to the chiral dimensions in order
to construct effective Lagrangians. 
We have shown that chiral dimensions simply count the loop order
of diagrams, and in that sense they have a topological nature.
Our approach is also consistent with the basic philosophy
of NDA and shows how the simple NDA rules need to be generalized.

The discussion of power counting presented here provides a
general and unified framework for constructing low-energy effective
theories of a strong sector. It encompasses chiral perturbation
theory weakly coupled to gauge fields, the electroweak chiral Lagrangian
and further theories of this kind with other patterns of symmetries
and symmetry breaking.

\section*{Acknowledgements}

We thank Rodrigo Alonso, Gino Isidori, Marc Knecht and 
Aneesh Manohar for a careful reading of the manuscript and for
useful remarks and suggestions. 
This work was performed in the context of the ERC Advanced Grant
project `FLAVOUR' (267104) and was supported in part by the 
DFG cluster of excellence `Origin and Structure of the Universe'.


\end{document}